# Biochemically Altered Human Erythrocytes as a Carrier for Targeted Delivery of Primaquine: an *In Vitro* Study

Fars K. Alanazi[1,2], Gamal El-Din I. Harisa[1], Ahmad Maqboul[1], Magdi Abdel-Hamid[1], Steven H. Neau[3], and Ibrahim A. Alsarra[2,4]

[1]Kayyali Chair for Pharmaceutical Industry, College of Pharmacy, King Saud University, P. O. Box 2457, Riyadh 11451, Saudi Arabia, [2]Department of Pharmaceutics, College of Pharmacy, King Saud University, P. O. Box 2457, Riyadh 11451, Saudi Arabia, [3]Department of Pharmaceutical Sciences, Philadelphia College of Pharmacy, University of the Sciences in Philadelphia, 600 South 43rd Street, Philadelphia, PA 19104, USA, and [4]Center of Excellence in Biotechnology Research, King Saud University, P. O. Box 2460, Riyadh 11451, Saudi Arabia

The aim of this study was to investigate human erythrocytes as a carrier for targeted drug delivery of primaquine (PQ). The process of PQ loading in human erythrocytes, as well as the effect of PQ loading on the oxidative status of erythrocytes, was also studied. At PQ concentrations of 2, 4, 6, and 8 mg/mL and an incubation time of 2 h, the ratios of the concentrations of PQ entrapped in erythrocytes to that in the incubation medium were 0.515, 0.688, 0.697 and 0.788, respectively. The maximal decline of erythrocyte reduced glutathione content was observed at 8 mg/mL of PQ compared with native erythrocytes $p < 0.001$. In contrast, malondialdehyde and protein carbonyl were significantly increased in cells loaded with PQ ($p < 0.001$). Furthermore, osmotic fragility of PQ carrier erythrocytes was increased in comparison with unloaded cells. Electron microscopy revealed spherocyte formation with PQ carrier erythrocytes. PQ-loaded cells showed sustained drug release over a 48 h period. Erythrocytes were loaded with PQ successfully, but there were some biochemical as well as physiological changes that resulted from the effect of PQ on the oxidative status of drug-loaded erythrocytes. These changes may result in favorable targeting of PQ-loaded cells to reticulo-endothelial organs. The relative impact of these changes remains to be explored in ongoing animal studies.

**Key words:** Primaquine, Carrier erythrocytes, Osmotic fragility, Glutathione, Malondialdehyde, Protein carbonyl

## INTRODUCTION

Primaquine (PQ), an 8-aminoquinoline, is widely used for the treatment of malarial infection, but its therapeutic use has been limited because of its side effects, such as gastrointestinal disturbances, granulocytopenia and hemolytic anemia (Kaur et al., 2010). In addition, PQ has a short plasma half-life, necessitating frequent administration, which then aggravates its adverse effects. Moreover, its adverse effects are enhanced because the drug must be frequently administered at high doses, due to its limited oral bioavailability (Vale et al., 2009). To reduce PQ side effects and increase its therapeutic efficacy, several studies have suggested different formulations for PQ administration such as transdermal systems (Mayorga et al., 1997), nanoparticles (Bhadra et al., 2005), microparticles (Nishi and Jayakrishnan, 2004), and PQ-loaded erythrocytes (Talwar and Jain, 1992). Pharmacokinetic studies revealed that these formulations achieved sustained release (Vale et al., 2009), dose reduction (Rodrigues et al., 1995), improved stability (Stensrud et al., 2000; Singh and Vingkar, 2008), and enhanced liver delivery (Dierling and Cui, 2005) when compared to that of free PQ administration.

Correspondence to: Ibrahim A. Alsarra, Department of Pharmaceutics, College of Pharmacy, King Saud University, Riyadh 11451, Saudi Arabia
Tel: 966-1-467-7504, Fax: 966-1-467-6363
E-mail: ialsarra@ksu.edu.sa

Erythrocytes are preferred as a drug delivery system because they are biocompatible and biodegradable, have a long half-life, and can be loaded with a variety of drugs (Patel et al., 2008). Anticancer, antiviral, and antiparasitic drugs are examples of therapeutic agents that have been loaded into erythrocytes (Gutiérrez Millán et al., 2004). Erythrocytes were used either as a carrier for sustained release of the drugs or to accomplish targeted delivery of the drugs to infected organs (Jaitely et al., 1996; Hamidi et al., 2007a).

The safety and utilization of erythrocytes as carrier systems have been previously discussed (Adams et al., 2003). The major problem encountered in the use of biodegradable natural cells as drug carriers is that they are removed *in vivo* by the reticuloendothelial system (RES) as a result of modifications that occur during the loading procedure in cells. Although this expands the capability of erythrocytes to target the RES, it seriously limits their life-span as long-circulating drug carriers in circulation and, in some cases, may pose toxicological problems (Hamidi et al., 2007a; Patel et al., 2008).

Utilization of erythrocytes as a drug carrier in humans also has the inherited problems of transfusion of blood from one to another, possible contamination due to the origin of the blood, the equipment used, and the loading environment (Adams et al., 2003). Therefore, screenings of these carriers for the absence of diseases as well as rigorous controls are required for the collection and handling of erythrocytes to eliminate any risk of contamination (Jain and Jain, 1997; Valbonesi et al., 2001).

Also, erythrocytes as a drug carrier raises other potential concerns due to the changes in their biochemical nature.

Drugs or other bioactive agents can be loaded in erythrocytes either by physical methods (*e.g.,* an electrical pulse method or osmosis-based systems) or by chemical methods (*e.g.,* chemical perturbation of the erythrocyte membrane) (Gopal et al., 2007). For successful entrapment into the erythrocytes, the drug should have a degree of water solubility and resistance to degradation within erythrocytes (Hamidi et al., 2007b). Certain drugs have been entrapped in erythrocytes by endocytosis, including vinblastine, chlorpromazine, hydrocortisone, propranolol, tetracaine, retinol, and PQ (Gopal et al., 2007).

Many drugs and chemicals are able to react with membrane lipids and proteins through crosslinking or oxidation, which may alter the asymmetry of the erythrocyte membrane. PQ, for example, may expose erythrocytes to excess oxidative stress, leading to depletion of the major antioxidant (reduced glutathione, GSH) and an increase in membrane lipid peroxidation and liberation of malondialdehyde (MDA) as a peroxidation product (Becker et al., 2004). Oxidative injury to erythrocytes leads to membrane asymmetry, which increases the possibility of erythrocyte uptake by macrophages, and therefore accelerates their elimination from the circulation by the RES (Zwaal and Schroit, 1997). An increase in the amount of oxidatively modified proteins is a hallmark of the ageing of erythrocytes (Robaszkiewicz et al., 2008). In the present study, the effect of PQ concentration and incubation time on the efficiency of PQ loading into human erythrocytes were determined. Furthermore, PQ-induced oxidative stress on erythrocytes was demonstrated by determination of GSH and MDA, whereas erythrocyte protein carbonyl content was measured as an indicator of protein oxidation. Also, the osmotic fragility of both native and PQ-loaded erythrocytes was determined to investigate the physiological state of erythrocyte membranes.

## MATERIALS AND METHODS

### Materials

Primaquine diphosphate, adenosine triphosphate, Ellman's reagent [5,5-dithio-bis(2-nitrobenzoic acid)], thiobarbituric acid, 1,1,3,3-tetraethoxypropane and GSH were purchased from Sigma-Aldrich Chemical Company; glutaraldehyde and HPLC grade acetonitrile and methanol were acquired from Merck & Co., Inc. (Whitehouse Station). All other chemicals used were of analytical grade.

### Instrumentation

In these investigations we used the following instruments: a Coulter® LH 780 hematology analyzer (Beckman Coulter, Inc.); a Spectro UV-Vis Split Beam PC, model UVS-2800 (Labomed, Inc.); a high performance liquid chromatography system consisting of a Waters 1525 binary pump, a Waters 717 Plus autosampler, a Waters 2487 dual wavelength absorbance detector (Waters Inc.), a Nucleodur® $C_{18}$ Gravity column (5 μm, 4.6 × 250 mm) (Macherey-Nagel GmbH & Co. KG); and a CT5 centrifuge.

### Blood collection and preparation

This study was approved by the research center ethics committee of King Saud University, College of Pharmacy, Riyadh, Saudi Arabia and informed consent was obtained from each of the blood donors. The blood samples were used immediately after collection. They were collected in heparinized vacutainers from healthy individual blood donors and centrifuged for 5 min at 5000 rpm. The plasma and the buffy coat were removed by aspiration. The erythrocytes were washed three times in cold phosphate buffered saline (PBS) with centrifugation for 5 min at 5000 rpm (Hamidi et al., 2007b).

**PQ loading procedures**

Erythrocyte suspensions with hematocrits adjusted to 40% were incubated with different concentrations of PQ (2, 4, 6, or 8 mg/mL) in PBS for 2 h at 37°C with mild agitation. Five hundred μL aliquots were taken at 30, 60, and 120 min, immediately centrifuged at 5000 rpm for 5 min, and used for extraction and quantification of PQ (Talwar and Jain, 1992).

**Erythrocyte counts**

Normal, PBS-suspended, and PQ-loaded erythrocytes were counted. The mean corpuscular volume (MCV), the mean corpuscular hemoglobin (MCH), and the mean corpuscular hemoglobin concentration (MCHC) were measured using a Coulter® LH 780 hematology analyzer.

**Scanning electron microscopy (SEM)**

A JEOL JSM-6380 LA scanning electron microscope (Jeol Ltd.) equipped with a digital camera, at a 20 kV accelerating voltage was used to evaluate morphological differences between normal and PQ-loaded erythrocytes. Both normal and 8 mg/mL PQ-loaded erythrocyte samples were processed as follows. After the samples were fixed in buffered glutaraldehyde, the aldehyde medium was drained off. The cells were rinsed 3 times for 5 min in phosphate buffer and post-fixed in osmium tetroxide for 1 h. The samples were then rinsed with distilled water and dehydrated using a graded ethanol series: 25, 50, 75, 100, and another 100%, each for 10 min. The samples were rinsed in water, removed, mounted on studs, sputter-coated with gold, and then viewed using SEM (Hamidi et al., 2007b).

**Determination of osmotic fragility**

A 25 μL erythrocyte sample was added to each of a series of 2.5 mL saline solutions containing 0.0 to 0.9 g % NaCl. After gentle mixing and standing for 15 min at room temperature, the erythrocyte suspensions were centrifuged at 5000 rpm for 5 min. The absorbance of the supernatant was measured at 540 nm (Kraus et al., 1997). The released hemoglobin was expressed as percentage absorbance of each sample in reference to a completely lysed sample prepared by diluting packed cells of each type with 1.5 mL of distilled water.

***In vitro* release of hemoglobin and PQ from carrier erythrocytes**

To study the *in vitro* release of both hemoglobin and PQ, carrier erythrocytes were resuspended in PBS to a final hematocrit of 30%. The suspensions were incubated at 37°C with gentle shaking and aliquots were removed at 0.5, 1, 2, 24, and 48 h. The aliquots were centrifuged at 5000 rpm for 5 min and the supernatant separated. The concentration of hemoglobin and PQ released into the PBS was determined. The release studies were performed in triplicate with carrier erythrocytes prepared using 2 h of incubation time and an initial PQ concentration of 8 mg/mL (Talwar and Jain, 1992; Hamidi et al., 2007b; Gutiérrez-Millán et al., 2008).

**HPLC analysis of PQ**

The PQ concentration was determined using a modified version of literature HPLC techniques (Talwar and Jain, 1992; Mayorga et al., 1997). Briefly, erythrocytes loaded with PQ were hemolysed by adding distilled water (1:1). The lysate was deproteinized with methanol (1:1.5). A 200 μL sample was mixed with 300 μL of methanol, vortexed for 30 sec, and then centrifuged for 15 min at 13,000 rpm. The supernatant was filtered and injected into the HPLC system. The mobile phase consisted of 0.05% trifluoroacetic acid and acetonitrile (75:25), with a 2 mL/min flow rate. The absorbance was measured at 265 nm.

**Determination of erythrocyte GSH content**

A literature spectrophotometric assay (Ko et al., 2008) was used for determination of GSH levels in erythrocyte samples. The peptide was precipitated from erythrocyte lysates by trichloroacetic acid (TCA). The mixture was filtered and 200 μL of the filtrate was added to 800 μL of 0.3 M phosphate buffer and 100 μL of 0.2% (w/v) Ellman's reagent for color development. The absorbance of the yellow color was measured at 412 nm. The concentration of GSH was calculated using a standard curve for GSH.

**Determination of erythrocyte lipid peroxidation**

Lipid peroxidation was assayed by MDA measurement using a published spectrophotometric method (Ohkawa et al., 1979). A mixture of 200 μL of 8% sodium dodecyl sulfate, 200 μL of 0.9% thiobarbituric acid, and 1.5 mL of 20% acetic acid were added to a 200 μL red blood cell lysate sample; then 1.9 mL of distilled water brought the volume to 4 mL. After boiling for 1 h, the mixture was cooled and 5 mL of an *n*-butanol and pyridine (15:1) solution was added. The mixture was centrifuged at 5000 rpm for 15 min and the absorbance was measured at 532 nm. Quantification of MDA levels was performed using tetraethoxypropane, the unstable malondialdehyde tetraethylacetal, as a standard.

### Assessment of erythrocyte protein oxidation

Protein carbonyl was assayed as a marker for erythrocyte protein oxidation, according to the method of Levine et al. (1994). Red blood cells were hemolysed, and proteins were precipitated by the addition of 10% TCA. The proteins were resuspended in 1.0 mL of 2 M HCl for the blank, and 2 M HCl containing 2% 2,4-dinitrophenyl hydrazine for test samples. After incubation for 1 h at 37°C, protein samples were washed with alcohol and ethyl acetate, and re-precipitated by addition of 10% TCA. The precipitated protein was dissolved in 6 M guanidine hydrochloride solution and absorbance was measured at 370 nm. Calculations were made using the molar extinction coefficient of $22 \times 10^3$ $M^{-1} \cdot cm^{-1}$ and the results are expressed as nmol carbonyls formed per mg protein. Total protein in the red blood cell (RBC) pellet was assayed according to the method of Lowry et al. (1951) using bovine serum albumin as the standard.

### Statistical analysis

The significance of the differences between native and loaded erythrocytes were analyzed by one way ANOVA followed by the Tukey Kramer multiple comparison test, using GraphPad® Prism Software, v. 5.01 (GraphPad Software, Inc.). Results with $p < 0.05$ were considered statistically significant.

## RESULTS AND DISCUSSION

Our intent was to use loaded erythrocytes as a drug delivery system to target delivery of PQ to the RES, in particular the liver and spleen. An elevated loading of PQ into erythrocytes was associated with a higher concentration of PQ in the incubation medium, in the 2 - 8 mg/mL concentration range (Table I). The highest level of loaded PQ was attained using 8 mg/mL of PQ with a 2 h incubation time.

Studies have shown that PQ can be loaded into erythrocytes by an endocytotic mechanism (Gopal et al., 2007; Hamidi et al., 2007a). Talwar and Jain (1992) characterized and evaluated the use of rat erythrocytes as carriers for PQ. In our study, we evaluated the extent of loading over time and studied the effect of PQ loading on biochemical changes in human erythrocytes.

**Table I.** Concentrations of loaded primaquine (mg/mL) in human carrier erythrocytes as a function of endocytosis time (min) and initial PQ concentration

| Loading time | Initial PQ concentration | Loaded PQ |
|---|---|---|
| 30 | 2 | 0.57 ± 0.07 |
|    | 4 | 1.60 ± 0.12 |
|    | 6 | 2.70 ± 0.13 |
|    | 8 | 4.30 ± 0.03 |
| 60 | 2 | 0.67 ± 0.05 |
|    | 4 | 2.00 ± 0.05 |
|    | 6 | 3.18 ± 0.12 |
|    | 8 | 4.99 ± 0.22 |
| 120 | 2 | 1.03 ± 0.19 |
|    | 4 | 2.75 ± 0.07 |
|    | 6 | 4.18 ± 0.27 |
|    | 8 | 6.30 ± 0.27 |

Data are expressed as mean ± S.E.M. (n = 6).

Endocytosis can be increased by certain conditions or the presence of certain factors, such as hypotonicity, EDTA, trypsin, and an energy source (Schrier et al., 1986). Erythrocyte energetics and calcium ions provide an opportunity for membrane invagination and fusion with formation of endocytotic vacuoles (Schrier et al., 1978). PQ-induced endocytosis is dependent on the persistence of erythrocyte energy sources (Matovcik et al., 1985). Our findings are in agreement with an earlier report that PQ loading is dependent on the concentration of drug in the surrounding media (Matovcik et al., 1985). Also, scanning of loaded erythrocytes using electron microscopy revealed the presence of pinocytotic vacuoles (Matovcik et al., 1985). Persistence of endocytotic vacuoles provides confirmation of entrapment of PQ. Matovcik et al. (1985) reported that membrane internalization associated with vacuole formation can be induced in intact human erythrocytes by incubating them with membrane active agents. Stomatocytosis and increased calcium membrane association can be produced by endocytosis-inducing drugs. A drug-induced increase in calcium membrane association may be a consequence of stomatocytosis and may secondarily be related to subsequent endocytosis. Moreover, PQ endocytosis is dependent on active metabolism manifested by a requirement for an energy source (Schrier et al., 1980).

Hematological parameters, such as MCV, MCH, and MCHC, were characterized. These parameters determine the influence of the encapsulation process on the hematological properties of the erythrocytes

**Table II.** Hematological parameters of control erythrocytes and loaded erythrocytes obtained with different concentrations of PQ (mg/mL)

| Parameters | Initial concentration of PQ | | | | |
| --- | --- | --- | --- | --- | --- |
| | Control | 2 | 4 | 6 | 8 |
| MCV (fl) | 84.8 ± 0.92 | 84.2 ± 1.07 | 88.4 ± 0.87 | 89.7 ± 1.23 | 97.5 ± 2.19*** |
| MCH (pg) | 29.3 ± 0.49 | 28.0 ± 0.69 | 29.4 ± 0.44 | 28.6 ± 0.72 | 27.0 ± 0.48* |
| MCHC (g/dL) | 33.7 ± 0.53 | 32.8 ± 0.34 | 33.5 ± 0.20 | 32.8 ± 0.22 | 30.7 ± 0.88*** |

MCV, mean corpuscular volume; MCH, mean corpuscular hemoglobin; MCHC, mean corpuscular hemoglobin concentration; fl, femtoliters. Data are expressed as mean ± S.E.M. (n = 6). *Significantly different at $p < 0.05$; ***Significantly different at $p < 0.001$ compared to unloaded erythrocytes (control).

(Gutiérrez-Millán et al., 2008). Table II presents the mean hematological parameters of the PQ-loaded erythrocytes obtained with different PQ concentrations and values for the same cells before the loading procedures (the control cells). A trend toward increasing MCV with an increase in the PQ concentration is evident, and one way ANOVA revealed that, compared with unloaded erythrocytes, the 8 mg/mL PQ-loaded cells have a significantly increased MCV ($p < 0.001$). This indicates that, in general, loaded RBCs will be larger in size than native cells, and this can be attributed to internalization of PQ from the surrounding medium. The MCH of loaded cells was lower than that of native cells by about 7.85% at the same PQ level, which could be due to hemoglobin leakage from loaded erythrocytes. There is a significant reduction of MCHC compared with native erythrocytes ($p < 0.001$). These results indicate that the loading procedure induced significant changes in the MCV in comparison with native cells, which implies retention of *in vitro* hematologic behaviors for PQ-loaded erythrocytes.

GSH is the main antioxidant in erythrocytes that protects proteins and lipids from oxidative damage. The oxidation of such molecules can result in loss of membrane integrity. Furthermore, GSH maintains sulfhydryl groups of proteins in the reduced state (Jain, 1984). Accordingly, it is relevant to investigate the susceptibility of erythrocytes to PQ-induced oxidative damage. In comparison to native erythrocytes, PQ-loading induced a significant reduction in erythrocyte GSH content (Fig. 1).

A significant elevation of MDA content for erythrocytes loaded with PQ in relation to unloaded erythrocytes was also found (Fig. 2). An increase in the incubation time resulted in augmentation of the MDA level and a GSH concentration decline (Fig. 3). These results are clear indicators of enhanced oxidative stress due to the presence of PQ.

Erythrocyte membrane lipids are the most important targets of intracellular peroxidation, resulting in cell membrane damage (Jain, 1984). The presence of a high content of polyunsaturated fatty acids makes them more sensitive to oxidative injury (Chiu et al, 1989). Mandal et al. (2002) reported that treatment of human erythrocytes with drugs induces oxidative modification that accelerates their clearance from the blood. This oxidative stress alters the asymmetry within the erythrocyte membrane, and rapid elimination of oxidized erythrocytes from the circulation is accelerated by alteration of the asymmetry (Jain, 1984). Another report stated that oxidatively modified erythrocytes are flagged for phagocytosis (Tyurina et al., 2000).

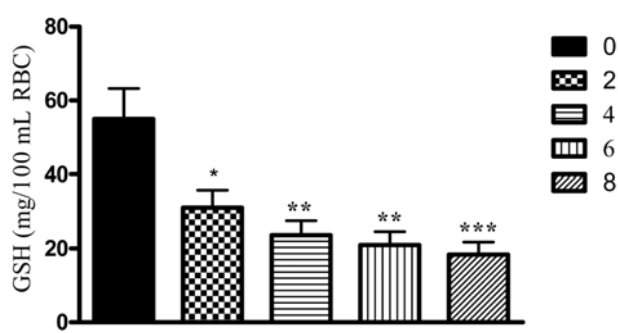

**Fig. 1.** Effect of different PQ (mg/mL) concentrations on erythrocyte glutathione content after 2 h incubation time. Data are expressed as mean ± S.E.M. (n = 5). *Significantly different at $p < 0.05$; **Significantly different at $p < 0.01$; ***Significantly different at $p < 0.001$, compared to unloaded erythrocytes (0 concentration).

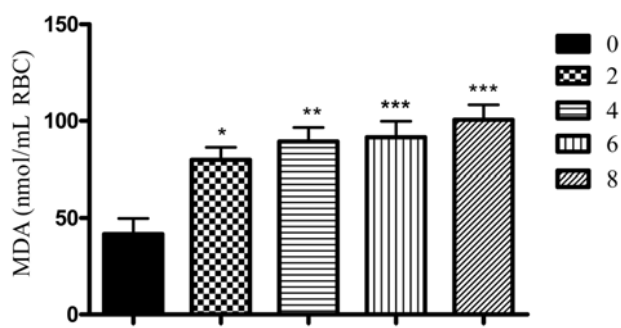

**Fig. 2.** Effect of different PQ (mg/mL) concentrations on erythrocyte malondialdehyde content after a 2 h incubation. Data are expressed as mean ± S.E.M. (n = 6). *Significantly different at $p < 0.05$; **Significantly different at $p < 0.01$; ***Significantly different at $p < 0.001$, compared to unloaded erythrocytes (0 concentration).

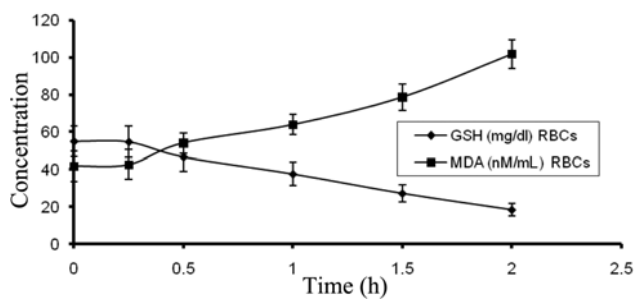

**Fig. 3.** Effect of PQ (8 mg/mL) on erythrocyte glutathione and malondialdehyde content as a function of incubation time. Data are expressed as mean ± S.E.M. (n = 6).

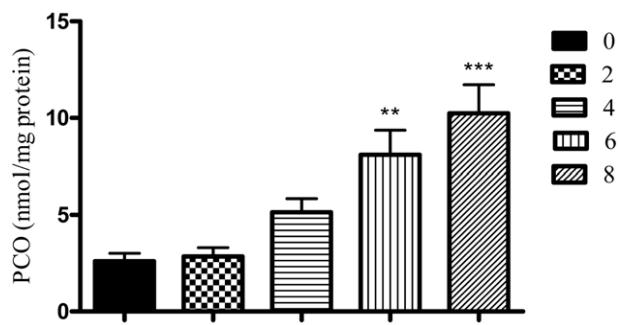

**Fig. 4.** Effect of different PQ (mg/mL) concentrations on erythrocyte protein carbonyl content after a 2 h incubation. Data are expressed as mean ± S.E.M. (n = 6). **Significantly different at $p < 0.01$; ***Significantly different at $p < 0.001$ compared to unloaded erythrocytes (0 concentration).

One of the functions of erythrocyte membrane proteins is the maintenance of aminophospholipids on the inner leaflet of the lipid bilayer. Moreover, membrane-bound proteinases serve as an antioxidant defense by removal of the oxidatively altered proteins (Beppu et al., 1994). The oxidative modification of proteins may be one of the factors responsible for the altered membrane asymmetry of oxidized RBC membranes (Dumaswala et al., 1997). Oxidative stress can induce denaturation of proteins through an increase in the formation of disulfide bonds and carbonyl groups (Rodrigues et al., 1995).

In the present work, we estimated protein carbonyl formation to assess PQ-mediated protein oxidation of loaded erythrocytes. We found a significant elevation of protein carbonyl (PCO) content of erythrocytes loaded with PQ after 2 h exposure to 6 and 8 mg/mL PQ (Fig. 4). These results are in agreement with a previous report that exposure of erythrocytes to chemicals increases membrane protein oxidation (Pandey et al., 2009). Dumaswala et al. (1997) reported that antioxidant membrane-bound proteinases are lost due to oxidative modification. Therefore, PQ-induced oxidation of protein, as indicated by an increase in PCO, is similar to effects reported under conditions of oxidation (Bukowska et al., 2008). It has also been reported that the increase in protein oxidation is due to oxidation of thiol groups and their subsequent formation of disulfide bonds (Robaszkiewicz et al., 2008). Thus, the induction of oxidative stress in carrier erythrocytes by PQ, either by decreasing the GSH content or by increasing lipid and protein oxidation, clearly occurs.

Osmotic fragility determines the susceptibility of erythrocytes to osmotic lysis (Vettore et al., 1984). Our investigation revealed that there is a significant increase in the osmotic fragility of loaded erythrocyte at 8 mg/mL PQ compared to unloaded erythrocytes (Table III). Our results are in agreement with those of Hamidi et al. (2001) who found that osmotic fragility of drug-loaded erythrocytes is increased compared with unloaded cells. Furthermore, another study found that osmotic fragility of carrier erythrocytes is higher than unloaded erythrocytes (Jain et al., 1997). The low resistance of loaded cells to osmotic changes

**Table III.** Erythrocyte osmotic fragility of unloaded erythrocytes and erythrocytes loaded with PQ (8 mg/mL). Values are percent hemolysis in corresponding salt concentrations

| NaCl (%) | Unloaded erythrocytes (%) | Erythrocytes loaded with PQ (%) |
|---|---|---|
| 0.1 | 96.8 ± 4.80 | 98.5 ± 5.45 |
| 0.2 | 89.6 ± 4.60 | 94.9 ± 4.58 |
| 0.3 | 84.4 ± 4.14 | 89.6 ± 3.17 |
| 0.4 | 74.1 ± 3.04 | 87.7 ± 2.47* |
| 0.5 | 29.7 ± 3.42 | 53.4 ± 4.48* |
| 0.6 | 12.3 ± 1.97 | 34.0 ± 3.53** |
| 0.7 | 7.35 ± 1.34 | 21.4 ± 2.39* |
| 0.8 | 3.02 ± 0.66 | 12.4 ± 1.73* |
| 0.9 | 0.95 ± 0.32 | 2.76 ± 0.42 |

Data are expressed as mean ± S.E.M. (n = 6). **Significant at $p < 0.001$; *Significant at $p < 0.01$ compared to unloaded erythrocytes (control).

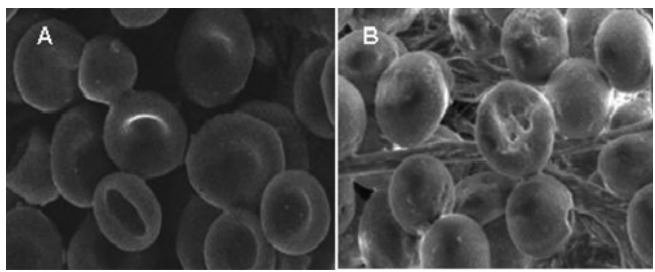

**Fig. 5.** Scanning electron microscopy of erythrocytes. (**A**) unloaded erythrocytes have a normal biconcave shape and (**B**) PQ (8 mg/mL)-loaded erythrocytes have a more spherical shape (Magnification 2000×).

may be due to the loss of erythrocyte membrane integrity. This may be caused by PQ-induced oxidative stress of erythrocyte membrane lipids and protein.

The main morphological change in PQ-loaded cells, as revealed by scanning electron microscopy, was the transformation of loaded cells from biconcave (normal) to near spherocytes, as illustrated in Fig. 5. The greatest change was observed with erythrocytes loaded with 8 mg/mL PQ. Attainment of a spherical shape due to drug loading makes the erythrocytes more fragile. The fragile cells may be destroyed and rapidly cleared from the circulation by macrophages (Talwar and Jain, 1992). Further studies are required to elucidate and demonstrate erythrophagocytosis of PQ-loaded erythrocytes.

Additionally, results from the present study demonstrated that carrier erythrocytes can be used as a sustained release delivery system for PQ. Erythrocytes exposed to 8 mg/mL PQ for 2 h were chosen for the *in vitro* release studies. Table IV presents the release pattern for PQ and hemoglobin from loaded erythrocytes to PBS over a 48 h period. Following an initial burst release over the first 2 h, the rate of release was reasonably consistent, although the drug release rate was 30% higher than that of hemoglobin. During the first 2 h period, about 55% of PQ was released from loaded cells to PBS, but only 10% of the hemoglobin. The drug apparently diffused readily

**Table IV.** Percent drug and hemoglobin release from PQ-loaded erythrocytes over time

| Time (h) | PQ (%) | Hemoglobin (%) |
|---|---|---|
| 0.5 | 29.3 ± 3.45 | 2.22 ± 0.42 |
| 1 | 40.7 ± 2.35 | 4.29 ± 0.63 |
| 2 | 55.0 ± 2.96 | 10.2 ± 1.84 |
| 24 | 63.8 ± 4.15 | 21.9 ± 4.11 |
| 48 | 76.4 ± 3.24 | 26.7 ± 2.26 |

Data are expressed as mean ± S.E.M. (n = 3).

since cell membrane lysis was not essential for release of PQ from loaded erythrocytes. The factors that determine drug release from carrier erythrocytes are size and the ionization of the drug molecule (Eichler et al., 1985). PQ diffused through the lipid bilayer to PBS, as shown with lipophilic drugs (Lewis and Alpar, 1984).

PQ entrapment by erythrocyte endocytosis provides a good yield when the loading medium is 8 mg/mL PQ. As a result of the PQ loading process, the erythrocyte GSH level is decreased, making membrane lipids and proteins more susceptible to oxidative modifications, as demonstrated by an increase in the MDA and PCO content. There is a change in erythrocyte morphology attributable to PQ loading. PQ-loaded cells exhibit sustained release of PQ over a period of 48 h.

## ACKNOWLEDGEMENTS

The authors would like to thank Dr. Omar Abdel-Kader, SEM Unit, Zoology Department, College of Science, King Saud University, for his assistance with the SEM analysis. This work was funded by the Centre of Excellence in Biotechnology Research (grant number CEBROF-1430/2), Ministry of Higher Education, Riyadh, Saudi Arabia.